\documentclass[english,pop,showpacs,reprint,superscriptaddress]{revtex4-1}
\usepackage[T1]{fontenc}
\usepackage[latin9]{luainputenc}
\usepackage{color}
\usepackage{amsmath}
\usepackage{amssymb}
\usepackage{graphicx}
\usepackage{wasysym}
\usepackage{soul}
\makeatletter
\usepackage{hyperref}

\makeatother

\usepackage{babel}
\begin{document}

\title{On the Rayleigh\textendash Kuo criterion for the tertiary instability of zonal flows}

\author{Hongxuan Zhu}

\affiliation{Department of Astrophysical Sciences, Princeton University, Princeton,
NJ, 08544 }

\affiliation{Princeton Plasma Physics Laboratory, Princeton, NJ 08543}

\author{Yao Zhou}

\affiliation{Princeton Plasma Physics Laboratory, Princeton, NJ 08543}

\author{I. Y. Dodin}

\affiliation{Department of Astrophysical Sciences, Princeton University, Princeton,
NJ, 08544 }

\affiliation{Princeton Plasma Physics Laboratory, Princeton, NJ 08543}
\begin{abstract}
This paper reports the stability conditions for intense zonal flows
(ZFs) and the growth rate $\gamma_{{\rm TI}}$ of the corresponding
``tertiary'' instability (TI) within the generalized Hasegawa\textendash Mima
plasma model. The analytic calculation extends and revises Kuo's analysis
of the mathematically similar barotropic vorticity equation for incompressible
neutral fluids on a rotating sphere [H.-L. Kuo, J. Meteor. $\textbf{6}$,
105 (1949)]; then, the results are applied to the plasma case. An
error in Kuo's original result is pointed out. An explicit analytic
formula for $\gamma_{{\rm TI}}$ is derived and compared with numerical
calculations. It is shown that, within the generalized Hasegawa--Mima model, a sinusoidal ZF is TI-unstable if and only if it satisfies the Rayleigh--Kuo criterion (known from geophysics) and that the ZF wave number exceeds the inverse ion sound radius. For non-sinusoidal ZFs, the results are qualitatively similar. As a corollary, there is no TI in the geometrical-optics
limit, i.e., when the perturbation wavelength is small compared to the ZF scale. 
This also means that the traditional wave kinetic equation, which is derived under the
geometrical-optics assumption, cannot adequately describe the ZF stability.
\end{abstract}
\maketitle

\section{Introduction}

\label{sec:section1}

Sheared plasma flows driven by turbulence, which are also known as
zonal flows (ZFs), significantly affect transport in magnetically
confined plasmas \cite{Diamond2005review,Fujisawa09,Lin98,Biglari90,Dorland00,Jenko00}.
Hence, they have been actively studied in literature. The linear stage
of the zonostrophic instability (ZI) that produces ZFs out of homogeneous
turbulence is now largely understood   \cite{Smolyakov2000a,Smolyakov2000b,Srinivasan2012,Parker2013,Parker2014,Farrell07,Farrell09,Constantinou16}.
 In contrast, the factors that limit the amplitudes of the nonlinear
ZFs have not been identified with certainty. The well-known predator-prey
model \cite{Malkov2001,Miki2012,diamond1994,kim2003} predicts that
the saturation amplitudes depend on the ZF collisional damping rates,
which are introduced in an $\textit{ad hoc}$ manner. However, the predator-prey
oscillations are also possible in the absence of dissipation  \cite{Zhu18a}, so the
ZF damping rate may not be the only important factor. One may wonder
then whether simple parameters can be identified that constrain the
ZF amplitude more robustly, i.e., without invoking $\textit{ad hoc}$
energy losses.

Here, we study a specific aspect of this problem, namely, the instability
of a prescribed ZF, which is known as the tertiary instability (TI)
\cite{Rogers2000,Rogers2005,kim2002,Numata2007,St-Onge2017,Singh2016,Rath18}.  The plasma is modeled within the generalized Hasegawa\textendash Mima equation (gHME), so our definition of the TI is different from that in Refs.~\cite{Rogers2000,Rogers2005}, where this instability was attributed to the ion-temperature gradient (absent in the gHME). However, our definition of the TI is similar to those in the majority of relevant papers \cite{kim2002,Numata2007,St-Onge2017,Singh2016}. 

As pointed out in Ref.~\cite{Parker2014phd}, the gHME is analogous
to the barotropic vorticity equation that describes neutral flows in the atmospheres of rotating planets  \cite{Kuo1949,Lipps1962}, where the Coriolis parameter
$\beta$ plays the same role as the density gradient in the gHME.
Therefore, the TI can be understood as analogous to the Kelvin\textendash Helmholtz
instability (KHI) \cite{Drazin1966,Drazin1981book} of a neutral flow
modified by nonzero $\beta$. (The analogy between the KHI and the
TI is also mentioned in Refs.~\cite{kim2002,Numata2007}.) The latter
modification was first studied by Kuo \cite{Kuo1949}, who generalized
the famous Rayleigh's inflection theorem for the KHI to the case of
nonzero $\beta$. The resulting criterion, known as the Rayleigh\textendash Kuo
(RK) necessary condition for instability, is widely cited in geophysics
literature but has not been popular in plasma physics literature with
only few exceptions \cite{Numata2007,Parker2016}. In Ref.~\cite{Numata2007},
it was mentioned that the criterion provides a good estimate for the
TI threshold, but no rigorous analysis was presented that would address
the necessary and sufficient conditions for the TI. This and the fact
that Kuo's results are not entirely accurate (see Sec.~\ref{sec:section3})
warrants a careful examination of the subject.

In this paper, we restate the RK criterion within the gHME model.
Specifically, we identify omissions in Kuo's original paper \cite{Kuo1949},
propose an explicit formula for the TI growth rate, and compare it
with numerical calculations. It is shown that, within the generalized Hasegawa--Mima model, a sinusoidal ZF is TI-unstable if and only if it satisfies the Rayleigh--Kuo criterion (known from geophysics) and that the ZF wave number exceeds the inverse ion sound radius. For non-sinusoidal ZFs, the results are qualitatively similar.
As a corollary, there is no TI in the geometrical-optics (GO) limit,
i.e., when the perturbation wavelength is small compared to the ZF scale. This
also means that the traditional wave kinetic equation \cite{Diamond2005review,Smolyakov2000a,Smolyakov2000b,Malkov2001,Miki2012,kim2002,Smolyakov1999,Li2018}
cannot adequately describe the ZF stability. In particular, the WKE-based
analysis of the TI in Ref.~\cite{kim2002} actually addresses a different
instability, namely, a branch of the ZI, as will be explained in Sec.~\ref{sec:section5}.

Note that, in order to produce a rigorous analytic theory, we simplify the problem by limiting our consideration to the strong-ZF case, namely, the case when the ambient turbulence is negligible and the ZF can be considered laminar. The more general case, when the ambient turbulence is allowed to affect the ZF stability, was studied using the stochastic structural stability theory in Refs.~\cite{Farrell07,Farrell09,Constantinou16} and using the equivalent second-order cumulant expansion theory in Refs.~\cite{Srinivasan2012,Parker2014}. In particular, Refs.~\cite{Parker2014,Farrell07,Farrell09,Constantinou16} reported that the ZF-turbulence system undergoes a structural instability even when the laminar ZF with the same amplitude would be RK-stable. Also, Ref.~\cite{Constantinou16} shows that the least-damped eigenmode changes its structure in the presence of the ambient turbulence and hence can become unstable. Correspondingly, the stability criterion that we report here can be considered as determining  the \textit{upper bound} of the ZF amplitude in a stable equilibrium. (Notably, the roles of stable eigenmodes is also discussed in Refs.~\cite{Pueschel16,Hatch16,Fraser17}.) 

Our paper is organized as follows. The basic equations are introduced
in Sec.~\ref{sec:section2}. In Sec.~\ref{sec:section3}, we obtain
the two necessary conditions and analytically calculate the TI growth
rate in the context that includes both geophysical and plasma settings
equally. The comparison between the analytic formula for the growth
rate and the numerically found eigenvalues is presented in Sec.~\ref{sec:section4}.
The ramifications of our theory that are specific to the plasma case
(as opposed to Kuo's geophysical problem) are discussed in Sec.~\ref{sec:section5}.
The generalization to non-sinusoidal ZF is discussed in Sec.~\ref{sec:section6}.
Our main conclusions are summarized in Sec.~\ref{sec:section7}.

\section{Basic equations}

\label{sec:section2}

\subsection{The generalized Hasegawa\textendash Mima equation}

\begin{figure}
\includegraphics[width=0.7\columnwidth]{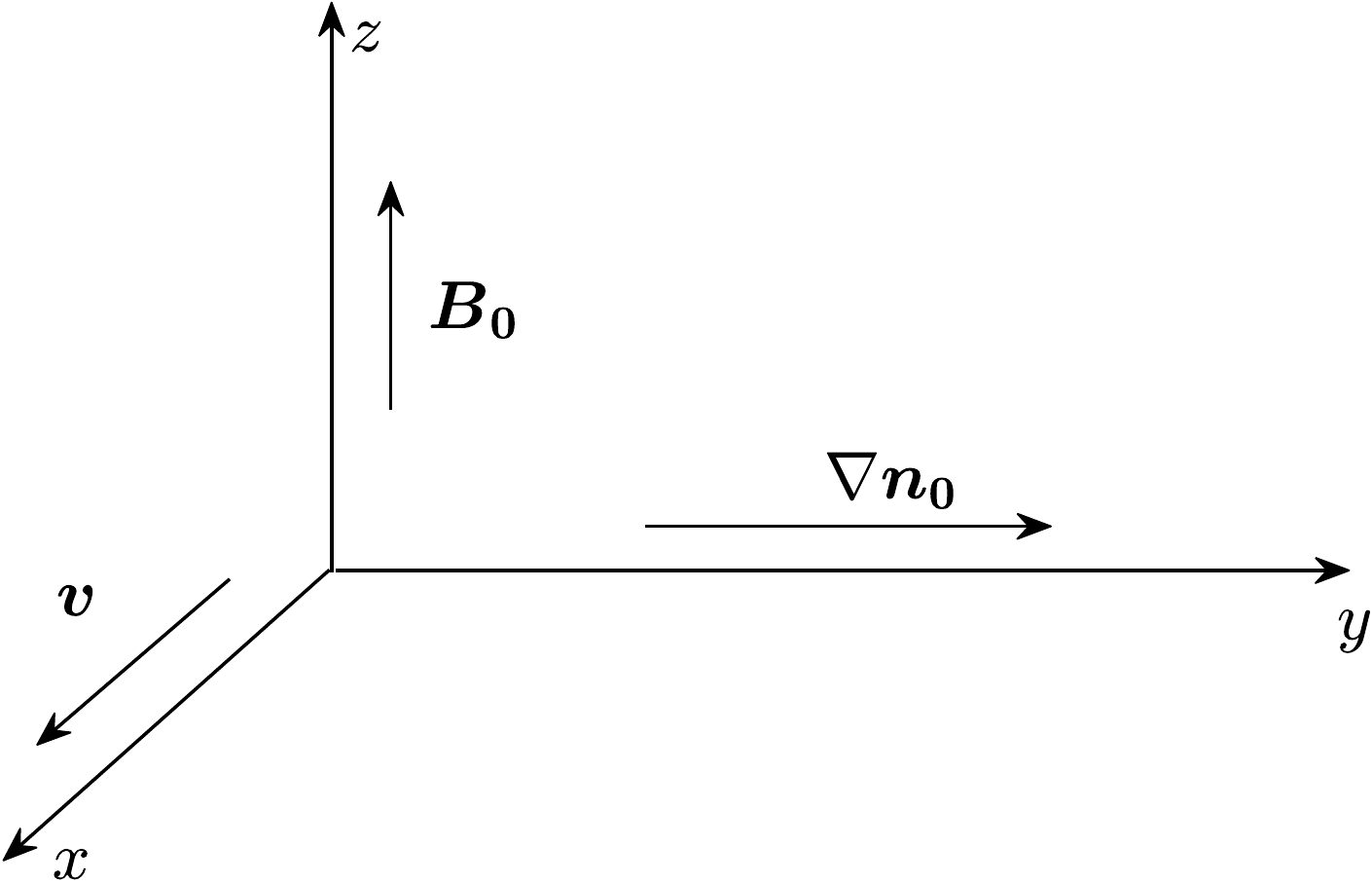}

\caption{The assumed coordinate system. Here, $\boldsymbol{B}_{0}$ is the
magnetic field, $n_{0}$ is the background electron density, and $\boldsymbol{v}$
is the ZF velocity.\label{fig:II_configuration}}
\end{figure}

First, let us introduce the original Hasegawa\textendash Mima equation
\cite{Hasegawa1977}. Consider a collisionless plasma in a uniform
magnetic field $\boldsymbol{B}_{0}$ in the $z$ direction, with the
equilibrium gradient of the background electron density $n_{0}$ in
the $y$ direction (Fig.~\ref{fig:II_configuration}). Ions are assumed
cold, while electrons are assumed to have a finite temperature $T_{e}$.
Suppose that perturbations to the electric field $\boldsymbol{E}$
are electrostatic, $\boldsymbol{E}=-\nabla\delta\varphi$, where $\delta\varphi(t,\boldsymbol{x})$
is the corresponding electrostatic potential on the two-dimensional
plane $\boldsymbol{x}\doteq(x,y)$. The electron response to $\boldsymbol{E}$
is adiabatic (yet see below), while the ion response can be described
by the $\boldsymbol{E}\times\boldsymbol{B}_{0}$ drift and the polarization
drift. Then, assuming the quasi-neutrality condition, the evolution
of $\delta\varphi$ is described by the (original) Hasegawa--Mima equation
\begin{multline}
\frac{\partial}{\partial t}\left[(\rho_{s}^{2}\nabla^{2}-1)\delta\varphi\right]\\
+\boldsymbol{u}_{E}\cdot\nabla\left[(\rho_{s}^{2}\nabla^{2}-1)\delta\varphi\right]+V_{*}\,\frac{\partial\delta\varphi}{\partial x}=0.\label{eq:II_hme_pre}
\end{multline}
Here, $\rho_{s}\doteq c_{s}/\Omega_{i}$ is the ion sound radius (we
use $\doteq$ to denote definitions), $c_{s}\doteq\sqrt{ZT_{e}/m_{i}}$
is the ion sound speed, $Z$ is the ion charge number, $\Omega_{i}\doteq Z|e|B_{0}/m_{i}$
is the ion gyrofrequency, $e$ is the electron charge, $\boldsymbol{u}_{E}\doteq\hat{\boldsymbol{z}}\times\nabla\delta\varphi/B_{0}$
is the $\boldsymbol{E}\times\boldsymbol{B}_{0}$ velocity, $\hat{\boldsymbol{z}}$
is the unit vector along the $z$ axis, $V_{*}\doteq T_{e}/(L_{n}B_{0}|e|)$
is the electron diamagnetic drift velocity, and $L_{n}\doteq(-\partial\ln n_{0}/\partial y)^{-1}$
is the characteristic length scale of $n_{0}$. Also, $\nabla^{2}\doteq\partial^{2}/\partial x^{2}+\partial^{2}/\partial y^{2}$
is the Laplacian.

Let us measure time in units $1/\Omega_{i}$ and length in units $\rho_{s}$.
Let us also introduce a normalized potential $\varphi\doteq e\delta\varphi/T_{e}$
and a normalized ``generalized vorticity'' $w\doteq(\nabla^{2}-1)\varphi$.
Then, Eq.~(\ref{eq:II_hme_pre}) can be written in the following
dimensionless form:
\begin{equation}
\frac{\partial w}{\partial t}+(\hat{\boldsymbol{z}}\times\nabla\varphi)\cdot\nabla w+\beta\,\frac{\partial\varphi}{\partial x}=0,\label{eq:II_hme}
\end{equation}
where $\beta\doteq V_{*}/c_{s}$ is treated as a (positive) constant. Let us also introduce the zonal average as $\langle f\rangle\doteq\int_{0}^{L_{x}}fdx/L_{x}$,
where $L_{x}$ is the system length in the $x$ direction. Then, perturbations
governed by Eq.~(\ref{eq:II_hme_pre}) include ZFs and DWs. The former
are identified as zonal-averaged perturbations, and the latter are
identified as fluctuations with zero zonal average. Strictly speaking,
electrons respond differently to ZFs and drift waves (DWs). Specifically, the above
model can be made more realistic if one rewrites the governing equations
as follows: 
\begin{gather}
\frac{\partial w}{\partial t}+(\hat{\boldsymbol{z}}\times\nabla\varphi)\cdot\nabla w+\beta\,\frac{\partial\varphi}{\partial x}=0,\label{eq:II_ghme}\\
w=(\nabla^{2}-\hat{a})\varphi,\label{eq:II_ghme2}
\end{gather}
where $\hat{a}$ is an operator such that $\hat{a}=1$ for DWs and
$\hat{a}=0$ for ZFs \cite{Dorland1993Thesis,Hammett1993}. Equations
(\ref{eq:II_ghme}) and (\ref{eq:II_ghme2}) constitute the so-called
gHME \cite{Krommes2000}. Below, we use this model to study the TI.

\subsection{Reduction to an eigenvalue problem}

\label{subsec:section2B}

Consider a stationary ZF with $\varphi\text{=\ensuremath{\bar{\varphi}}}(y)$
and $w=\bar{\varphi}''$. (Hereafter, the prime denotes the derivative
with respect to $y$.) The ZF velocity is $\boldsymbol{v}=U(y)\hat{\boldsymbol{x}}$,
where $U(y)\doteq-\bar{\varphi}'$ and $\hat{\boldsymbol{x}}$ is
the unit vector in the $x$ direction. Consider a DW perturbation
$\tilde{\varphi}\doteq\varphi-\bar{\varphi}$ to this ZF. As mentioned
in Sec.~\ref{sec:section1}, we focus on the case when the initial
state is non-turbulent.  Then, $\tilde{\varphi}$
is small and can be described by the linearized Eqs.~(\ref{eq:II_ghme})
and (\ref{eq:II_ghme2}), namely,
\begin{gather}
\frac{\partial\tilde{w}}{\partial t}+U\frac{\partial\tilde{w}}{\partial x}+(\beta-U'')\frac{\partial\tilde{\varphi}}{\partial x}=0,\\
\tilde{w}=(\nabla^{2}-1)\tilde{\varphi}.
\end{gather}

Let us search for a solution in the form $\tilde{\varphi}=\phi(y)\exp(ik_{x}x-i\omega t)$
with constant $k_{x}$ and $\omega$. Then $\tilde{w}=(\partial^{2}/\partial y^{2}-k_{x}^{2}-1)\tilde{\varphi},$
and
\begin{equation}
\left(\frac{d^{2}}{dy^{2}}-\alpha^{2}-\frac{U''-\beta}{U-C}\right)\phi(y)=0,\label{eq:II_barotropic_vorticity}
\end{equation}
where $\alpha^{2}\doteq1+k_{x}^{2}$ and $C\doteq\omega/k_{x}$. This
equation is identical \cite{foot:coordinate} to the barotropic vorticity
equation studied by Kuo in Ref.~\cite{Kuo1949}. The only difference
is that in Kuo's case, $\alpha^{2}$ is not necessarily larger than
one. However, we will not make use of our specific expression for
$\alpha^{2}$ until Sec.~\ref{sec:section5}, where we will consider
the application of our results to the gHME explicitly. In this sense,
our main results are also applicable in the geophysical context. 

Equation (\ref{eq:II_barotropic_vorticity}) can be represented as
\begin{eqnarray}
A^{-1}(UA+\beta-U'')\phi & = & C\phi\label{eq:II_eigen}
\end{eqnarray}
($A\doteq d^{2}/dy^{2}-\alpha^{2}$), so it is understood as an eigenvalue
problem (with certain boundary conditions). Without loss of generality,
we assume $k_{x}>0$. Hence, if $\text{Im}\,C>0$, then $\omega\equiv k_{x}C$
has a positive imaginary part, which signifies an instability. 

Also note that the complex conjugate of Eq.~(\ref{eq:II_barotropic_vorticity})
is
\begin{equation}
\left(\frac{d^{2}}{dy^{2}}-\alpha^{2}-\frac{U''-\beta}{U-C^{*}}\right)\phi^{*}(y)=0.\label{eq:II_barotropic_complex_conjugate}
\end{equation}
This shows that if $\phi$ is a solution corresponding to a certain
phase velocity $C$, then $\phi^{*}$ is also a solution corresponding
to $C^{*}$. Hence, an instability is possible whenever $\text{Im}\,C$
is nonzero, because a mode with $\text{Im}\,C$ of any sign is always
accompanied by a mode with $\text{Im}\,C$ of the opposite sign. {[}Notably,
this is not the case at nonzero viscosity, because then Eq.~(\ref{eq:II_barotropic_vorticity})
acquires complex coefficients.{]}

\subsection{Floquet analysis}

For simplicity, let us assume (until Sec.~\ref{sec:section6}) an unbounded sinusoidal ZF profile,
\begin{equation}
U=u_{0}\cos qy.\label{eq:II_U_profile}
\end{equation}
(This periodic profile, although different from those typically used
in studies of neutral fluids \cite{Kuo1949,Lipps1962,Drazin1966,Drazin1962,Howard1964}
with fixed boundaries, is often used in numerical or theoretical modeling
in plasma physics \cite{Numata2007,Kobayashi2012,Ruiz2016}.) For
clarity, we assume $u_{0}>0$ and $q>0$. Then, Eq.~(\ref{eq:II_barotropic_vorticity})
is an ordinary differential equation with periodic coefficients whose
period is $L=2\pi/q$. Then, any solution of Eq.~(\ref{eq:II_barotropic_vorticity})
is decomposable into Floquet modes of the form \cite{Yakubovich1975book}
\begin{equation}
\phi(y)=\psi(y)e^{i\bar{q}y},\label{eq:III_psi_definition}
\end{equation}
where $\psi(y)$ is periodic such that $\psi(y+L)=\psi(y)$ for any
$y$, and $\bar{q}$ is a constant. Assuming $\phi$ is bounded (as
it would be the case, for instance, at periodic boundary conditions),
$\bar{q}$ must be real. Without loss of generality, we limit the
value of $\bar{q}$ to the first Brillouin zone, i.e., $-q/2\leq\bar{q}<q/2$. 

In what follows, we explore conditions under which these restrictions
lead to complex $C$, i.e., an instability. Any such instability is
by definition considered a TI. 

\section{Instability onset}

\label{sec:section3}

\subsection{The Rayleigh\textendash Kuo criterion}

\label{subsec:section3a}

First, let us repeat the RK argument for completeness. Let us multiply
Eq.~(\ref{eq:II_barotropic_vorticity}) by $\phi^{*}$ and consider
the imaginary part of the resulting equation
\begin{equation}
\phi^{*}\phi''-\phi\phi^{*}\,''-\left(\frac{U''-\beta}{U-C}-\frac{U''-\beta}{U-C^{*}}\right)|\phi|^{2}=0.
\end{equation}
By integrating this over $y$ from $0$ to $L$, we obtain
\begin{multline}
\int_{0}^{L}\left(\frac{U''-\beta}{U-C}-\frac{U''-\beta}{U-C^{*}}\right)|\phi|^{2}dy\\
=\int_{0}^{L}(\phi^{*}\phi''-\phi\phi^{*}\,'')dy\\
=(\phi^{*}\phi'-\phi\phi^{*}\,')\left|_{0}^{L}\right.=0,
\end{multline}
where we used Eq.~(\ref{eq:III_psi_definition}) and the periodicity
of $\psi$. If $\text{Im}\,C\neq0$, then we obtain
\begin{equation}
\int_{0}^{L}\,\frac{U''-\beta}{|U-C|^{2}}\,|\phi|^{2}dy=0.\label{eq:III_RK}
\end{equation}
Hence, $U''-\beta$ must change its sign somewhere in the integration
domain; i.e., there must be a location where $U''=\beta$. (This location
can be understood as the point where the ``vorticity'' $Z$, defined
via $dZ/dy\doteq U''-\beta$, has an extremum \cite{Lin1955book}.)
This is the RK necessary condition for a ZF to be unstable \cite{Kuo1949}.
For the sinusoidal profile {[}Eq.~(\ref{eq:II_U_profile}){]}, the
existence of $U''=\beta$ means that 
\begin{equation}
\beta<q^{2}u_{0}.
\end{equation}

The RK criterion is a generalization of Rayleigh's famous inflection-point
theorem to the case of nonzero $\beta$. Another famous necessary
condition, the Fjørtoft's theorem \cite{Drazin1981book}, has also been generalized to barotropic and baroclinic flows \cite{Pedlosky82book}, and we briefly mention it here for the sake of completeness. Let us multiply Eq.~(\ref{eq:II_barotropic_vorticity})
by $\phi^{*}$ and integrate the result over $y$ from $0$ to $L$:
\begin{gather}
\int_{0}^{L}\phi^{*}\phi''dy-\alpha^{2}\int_{0}^{L}|\phi|^{2}dy=\int_{0}^{L}\,\frac{U''-\beta}{U-C}\,|\phi|^{2}dy\nonumber \\
=\int_{0}^{L}\,\frac{(U''-\beta)(U-C^{*})}{|U-C|^{2}}\,|\phi|^{2}dy.\label{eq:III_prepre_fojtoft}
\end{gather}
From Eq.~(\ref{eq:III_psi_definition}), we have
\begin{multline}
\int_{0}^{L}\phi^{*}\phi''dy=\left(\phi^{*}\phi'\right)|_{0}^{L}-\int_{0}^{L}|\phi'|^{2}dy\\
=\left(\psi^{*}\psi'+i\bar{q}|\psi|^{2}\right)|_{0}^{L}-\int_{0}^{L}|\phi'|^{2}dy\\
=-\int_{0}^{L}|\phi'|^{2}dy.
\end{multline}
Therefore the left-hand side of Eq.~(\ref{eq:III_prepre_fojtoft})
is 
\begin{gather}
-\int_{0}^{L}|\phi'|^{2}dy-\alpha^{2}\int_{0}^{L}|\phi|^{2}dy<0.\label{eq:III_pre_fojtoft}
\end{gather}
Let us multiply Eq.~(\ref{eq:III_RK}) with $C^{*}-U_{*}$, where
$U_{*}$ is the value of $U$ where $U''=\beta$, and add the resulting
equation to Eq.~(\ref{eq:III_prepre_fojtoft}). Using Eq.~(\ref{eq:III_pre_fojtoft}),
this leads to
\begin{equation}
\int_{0}^{L}\,\frac{(U''-\beta)(U-U_{*})}{|U-C|^{2}}\,|\phi|^{2}dy<0.
\end{equation}
This means that $(U''-\beta)(U-U_{*})$ must be negative somewhere,
which is the generalization of Fjørtoft's theorem in the case of nonzero
$\beta$. For the sinusoidal profile,  this criterion
is always satisfied if the RK criterion is satisfied.

\subsection{Neutral eigenmodes}

Equation (\ref{eq:II_barotropic_vorticity}) has a special class of
solutions called ``neutral modes'' that correspond to real $C$.
(Such perturbations neither grow nor decay in time.) Below, we consider
the case $-u_{0}\le C\le u_{0}$, which requires a special treatment
due to possible singularities at $U=C$. (The case when $C<-u_{0}$
or $C>u_{0}$ will be considered in Sec.~\ref{subsec:section3c}.)
Near a singularity $y=y_{s}$, two linearly independent solutions
$\phi_{1,2}$ of Eq.~(\ref{eq:II_barotropic_vorticity}) can be obtained
using Frobenius method \cite{Kuo1949} and have the following asymptotics:
\begin{gather}
\phi_{1}=(y-y_{s})+a_{2}(y-y_{s})^{2}+a_{3}(y-y_{s})^{3}+...,\\
\phi_{2}=1+b_{1}(y-y_{s})+b_{2}(y-y_{s})^{2}+...\nonumber \\
+G_{s}\phi_{1}\ln(y-y_{s}),
\end{gather}
where $G_{s}\doteq(U_{s}''-\beta)/U{}_{s}'$. (The subscript $s$
means that the corresponding functions are evaluated at $y=y_{s}$.)
The general solution can be written as 
\begin{equation}
\phi=c_{1}\phi_{1}+c_{2}\phi_{2}.
\end{equation}
Near the singularity, one has $\phi\approx c_{2}$, $\phi'\approx c_{2}G_{s}\ln(y-y_{s})$,
and $\phi''\approx c_{2}G_{s}/(y-y_{s})$. 

Note that the above two solutions are invalid if $U_{s}'=0$, but
those modes have infinite enstrophy and are physically irrelevant
\cite{foot:enstrophy}. Modes with nonzero $G_{s}$ are physically
irrelevant too for the same reason. The only exception is when $c_{2}=0$,
i.e., $\phi=c_{1}\phi_{1}=0$ at the singularity. However, this is
impossible, which is seen as follows. Let us compare Eq.~(\ref{eq:II_barotropic_vorticity})
with the following equation:
\begin{equation}
F''=\frac{U''}{U-C}\,F.
\end{equation}
Between two neighboring singularities, we have $C>U$ and $\alpha^{2}+\beta/(C-U)>0$,
hence $\phi$ oscillates slower than $F$ according to the Sturm comparison
theorem \cite{Kuo1949}. However, $F=U-C$ is a solution that is zero
at both singularities, which means that $\phi$ cannot be zero at
both singularities. This rules out the possibility that $c_{2}=0$. (Note that this conclusion relies on the specific profile we chose here. It may fail if there is only one singularity, e.g., for a monotonic $U$.)

From the above discussion, we conclude that only those neutral modes
that have $G_{s}=0$ need be considered. This corresponds to $C=C_{n}\doteq-\beta/q^{2}$
for a sinusoidal ZF. (The subscript $n$ stands for ``neutral''.)
Let us substitute $C_{n}$ into Eq\@.~(\ref{eq:II_barotropic_vorticity})
along with Eq.~(\ref{eq:III_psi_definition}). This gives \begin{align}
\psi''+2i\bar{q}\psi'-\bar{q}^2\psi&=\left(\alpha^{2}+\frac{U''-\beta}{U-C_{n}}\right)\psi\notag\\
&=\left(\alpha^{2}+\frac{-q^{2}u_{0}\cos qy-\beta}{u_{0}\cos qy+\beta/q^{2}}\right)\psi\notag\\
&=\left(\alpha^{2}-q^{2}\right)\psi.\label{eq:III_barotropic_psi}
\end{align}This is an ordinary differential equation with constant coefficients, so its
solutions can be searched in the form $\psi=\exp(i\lambda y)$. Then,
we obtain
\begin{equation}
(\lambda+\bar{q})^{2}=q^{2}-\alpha^{2}.\label{eq:III_lambda}
\end{equation}
Since $\psi$ is periodic in $y$, we require that $\lambda=mq$,
where $m$ can be any integer. Then, Eq.~(\ref{eq:III_lambda}) becomes
\begin{equation}
\alpha^{2}=q^{2}-(\bar{q}+mq)^{2}.
\end{equation}
For clarity, suppose $\bar{q}>0$. (The case $\bar{q}<0$ can be analyzed
similarly.) Then, $\alpha^{2}\geq0$ is possible only if $m=-1$ or
$m=0$. Therefore, we have two choices of $\alpha^{2}$, and the corresponding solutions are
\begin{gather}
\alpha_{n1}^{2}=q^{2}-(q-\bar{q})^{2},\quad\psi_{n1}=e^{-iqy},\label{eq:III_neutral_1}\\
\alpha_{n2}^{2}=q^{2}-\bar{q}{}^{2},\quad\psi_{n2}=1.\label{eq:III_neutral_2}
\end{gather}
Since $\bar{q}<q/2$, we have $0\leq\alpha_{n1}^{2}<\alpha_{n2}^{2}\leq q^{2}$.
This shows that, indeed, $C=C_{n}$ corresponds to neutral modes.
More precisely, two neutral modes correspond to the two choices of
$\alpha^{2}$ given by Eqs.~(\ref{eq:III_neutral_1}) and (\ref{eq:III_neutral_2}).

\subsection{Unstable eigenmodes}

\label{subsec:section3c}

Now, let us consider $\alpha^{2}=\alpha_{n}^{2}+\Delta\alpha^{2}$, where
$\Delta\alpha^{2}$ is a small perturbation to a neutral-mode solution
(and the subscript $n$ stands for $n1$ or $n2$). The corresponding
perturbed eigenvalues and eigenmodes are some $C=C_{n}+\Delta C$
and $\psi=\psi_{n}+\Delta\psi$. The neutral eigenmode $\psi_{n}$
satisfies Eq.~(\ref{eq:III_barotropic_psi}), while $\psi$ is governed
by \begin{equation}
\psi''+2i\bar{q}\psi'-\bar{q}^2\psi=\left(\alpha_{n}^{2}+\Delta\alpha^{2}+\frac{U''-\beta}{U-C_{n}-\Delta C}\right)\psi.\label{eq:III_perturbed_psi}
\end{equation} Let us multiply Eq.~(\ref{eq:III_perturbed_psi}) by $\psi_{n}^{*}$
and the complex conjugate of Eq.~(\ref{eq:III_barotropic_psi}) by
$\psi$. By subtracting one result from the other results, we obtain
\begin{multline}
\psi''\psi_{n}^{*}-\psi\psi_{n}^{*}\,''+2i\bar{q}(\psi\psi_{n}^{*}\,'+\psi'\psi_{n}^{*})\\
=\left(\Delta\alpha^{2}+\frac{U''-\beta}{U-C_{n}-\Delta C}-\frac{U''-\beta}{U-C_{n}}\right)\psi\psi_{n}^{*}.
\end{multline}
Let us integrate this equation over $y$ from $0$ to $L$. Using
the periodicity of $\psi_{n}$ and $\psi$, we obtain
\begin{multline}
\int_{0}^{L}\left(\Delta\alpha^{2}+\frac{U''-\beta}{U-C_{n}-\Delta C}-\frac{U''-\beta}{U-C_{n}}\right)\psi\psi_{n}^{*}dy\\
=\left(\psi'\psi_{n}^{*}-\psi\psi_{n}^{*}\,'+2i\bar{q}\psi\psi_{n}^{*}\right)\left|_{0}^{L}\right.=0.\label{eq:III_integration}
\end{multline}
In the above integral, the term in the bracket is at least of the
first order of $\Delta\alpha^{2}$ and $\Delta C$. Hence we can approximate
$\psi\psi_{n}^{*}$ with $|\psi_{n}|^{2}$. Therefore,
\begin{gather}
\Delta\alpha^{2}\approx\frac{-J(\Delta C)}{\mathcal{N}},\label{eq:III_dispersion}
\end{gather}
where $\ensuremath{\mathcal{N}\doteq\int_{0}^{L}|\psi_{n}|^{2}dy}$
and
\begin{multline}
J(\Delta C)\doteq\\
\int_{0}^{L}\left(\frac{U''-\beta}{U-C_{n}-\Delta C}-\frac{U''-\beta}{U-C_{n}}\right)|\psi_{n}|^{2}dy.
\end{multline}

The integrand is analytic in the whole integration domain if $\text{Im}(\Delta C)$
is nonzero. In Ref.~\cite{Kuo1949}, Kuo Taylor-expanded this integrand
in terms of $\Delta C$ for all $y$, and concluded that $\text{Im}(\Delta C)>0$
if $\Delta\alpha^{2}<0$ and $\text{Im}(\Delta C)<0$ if $\Delta\alpha^{2}>0$;
then, unstable eigenmodes exist only when $\Delta\alpha^{2}<0$. However,
$\Delta C$ cannot be considered small near $U=C_{n}$, hence the
Taylor expansion in  $\Delta C$ is invalid there. The correct
way to calculate $J$ is as follows. Since $J(\Delta C=0)=0$, to
the first order we have 
\begin{equation}
J(\Delta C)=J'(0)\Delta C.\label{eq:III_Jdc}
\end{equation}
In combination with Eq.~(\ref{eq:III_dispersion}), this gives that, to the first order in $\Delta C$, \begin{gather}
\Delta C=-\frac{\mathcal{N}\Delta\alpha^{2}}{J'(0)}.\label{eq:III_dc}
\end{gather} (Here, the prime denotes the derivative with respect to $\Delta C$, unlike in the rest of the text, where the prime denotes the derivative with respect to $y$.) 

If $C_{n}$ is not equal to $-\beta/q^{2}$ but lies out of the range
of $U$ (i.e., $C_{n}<-u_{0}$ or $C_{n}>u_{0}$), then $J'(0)$ is
real. Accordingly, Eq.~(\ref{eq:III_dc}) gives that $\Delta C$
is real, because $\Delta\alpha^{2}$ and $\mathcal{N}$ are real by
definition. In other words, perturbations to such neutral modes are
stable and thus do not need to be considered for our purposes. 

\begin{figure}
\includegraphics[width=1\columnwidth]{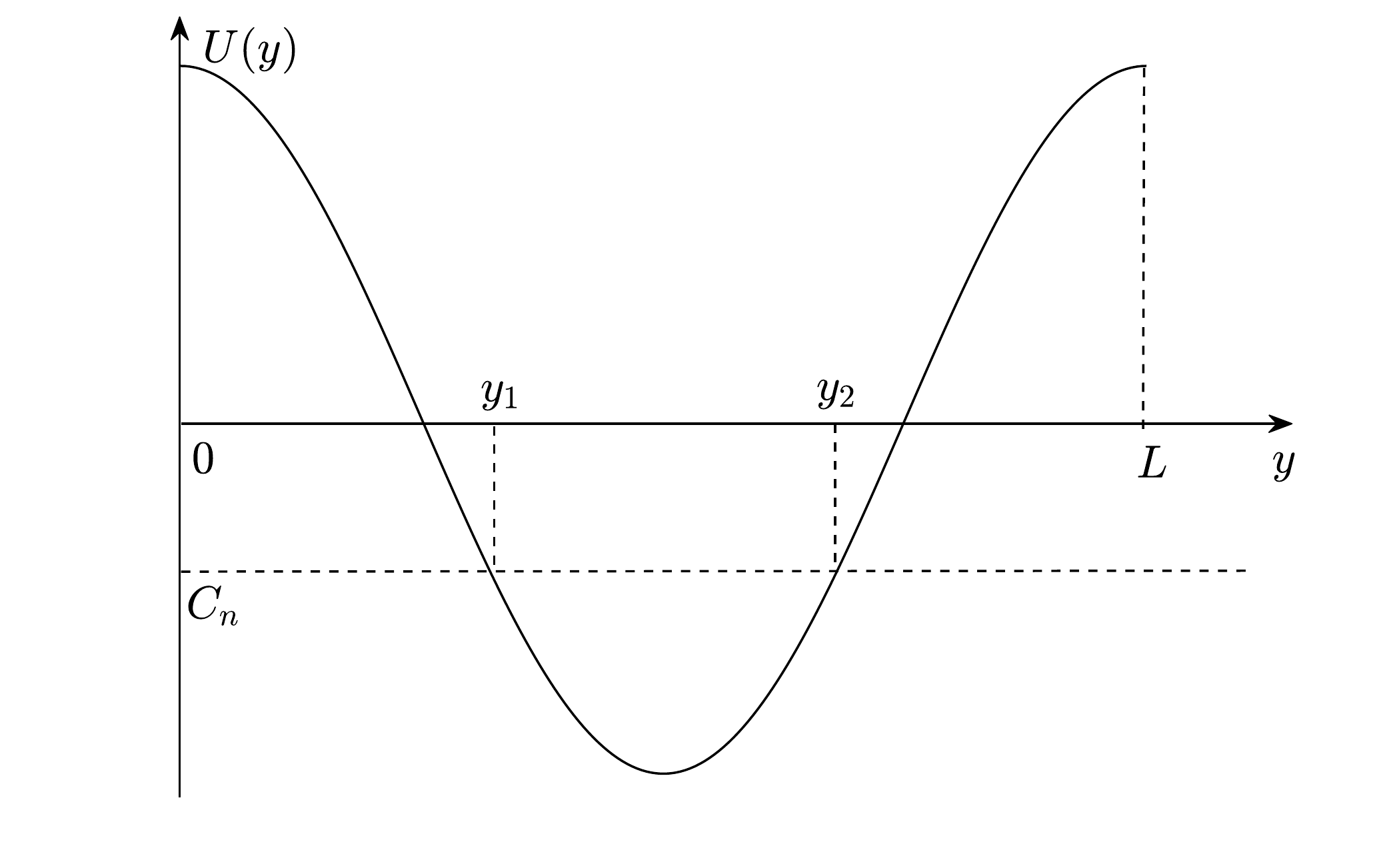}\caption{A schematic of the ZF profile $U(y)$ {[}Eq.~(\ref{eq:II_U_profile}){]}
and the locations of $y_{1}$ and $y_{2}$, where $U=C_{n}$. Note
that $U'(y_{1})<0$ and $U'(y_{2})>0$. \label{fig:III_illustration}}
\end{figure}

Now, let us consider $C_{n}=-\beta/q^{2}$. Using Eq.~(\ref{eq:III_neutral_1})
or (\ref{eq:III_neutral_2}), we have $|\psi_{n}|^{2}=1$, so $\mathcal{N}=L$.
Also, the expression for $J'(0)$ can be found as the limit of 
\begin{equation}
J'(\Delta C)=\int_{0}^{L}\,\frac{U''-\beta}{(U-C_{n}-\Delta C)^{2}}\,dy\label{eq:III_J'}
\end{equation}
at $\Delta C\to0$. Since the integrand has a first-order pole at
$U=C_{n}$ when $\Delta C=0$, it is convenient to use the Plemelj
formula,
\begin{equation}
\lim_{\epsilon\rightarrow0+}\frac{1}{x-i\epsilon}=\text{P.V.}\left(\frac{1}{x}\right)+i\pi\delta(x),
\end{equation}
where P.V. stands for principal value. Then, we have
\begin{multline}
\lim_{\Delta C\rightarrow0}\frac{U''-\beta}{(U-C_{n}-\Delta C)^{2}}=\text{P.V.}\frac{U''-\beta}{(U-C_{n})^{2}}\\
-i\pi\delta(y-y_{1})\cdot\left(\frac{U'''}{U'^{2}}\right)\bigg|_{y_{1}}\text{sgn}[\text{Im}(\Delta C)]\\
+i\pi\delta(y-y_{2})\cdot\left(\frac{U'''}{U'^{2}}\right)\bigg|_{y_{2}}\text{sgn}[\text{Im}(\Delta C)],
\end{multline}
where we used $U'|_{y_{1}}<0$ and $U'|_{y_{2}}>0$ (Fig.~\ref{fig:III_illustration}).
Hence, we obtain 
\begin{equation}
\lim_{\Delta C\rightarrow0}J'(\Delta C)=E-2iD\,\text{sgn}[\text{Im}(\Delta C)],
\end{equation}
where
\begin{multline}
E=\text{P.V.}\int_{0}^{L}\,\frac{U''-\beta}{(U-C_{n})^{2}}\,dy\\
=-\frac{q^{2}}{u_{0}}\,\text{P.V.}\int_{0}^{L}\,\frac{1}{\cos qy+\beta/q^{2}u_{0}}\,dy,
\end{multline}
and 
\begin{equation}
D=\pi\left(\frac{U'''}{U'^{2}}\right)\bigg|_{y_{1}}=-\pi\left(\frac{U'''}{U'^{2}}\right)\bigg|_{y_{2}}.
\end{equation}
A straightforward calculation gives
\begin{gather}
E=0,\\
D=\frac{\pi q}{u_{0}\sqrt{1-\varrho^{-2}}}>0,
\end{gather}
where $\varrho\doteq u_{0}q^{2}/\beta$ is what we call the Rayleigh\textendash Kuo
parameter \cite{Zhu18a}. Then, using Eq.~(\ref{eq:III_dc}) and
$L=2\pi/q$, we obtain  
\begin{equation}
\Delta C=\pm iu_{0}|\Delta\alpha^{2}|q^{-2}\,\sqrt{1-\varrho^{-2}}\label{eq:III_Im(C)}
\end{equation}
for $\Delta\alpha^{2}<0$. In contrast, if $\Delta\alpha^{2}>0$,
no self-consistent solution exists. This result is at variance with
Kuo's result but in agreement with our earlier observation (Sec.~\ref{sec:section2})
that, if $C$ is an eigenvalue, then so is $C^{*}$ \cite{foot:viscosity}. 

Since there are two neutral modes, the above calculation gives two
branches of unstable modes. Their growth rates are given by
$\gamma=|k_{x}{\rm Im}(\Delta C)|$, namely,
\begin{equation}
\gamma_{1,2}=|k_{x}u_{0}|q^{-2}G(\Delta\alpha_{1,2}^{2})\,\sqrt{1-\varrho^{-2}}.
\end{equation}
Here, $G(\Delta\alpha^{2})\doteq|\Delta\alpha^{2}|H(-\Delta\alpha^{2})$, 
where $H$ is the Heaviside step function, and
\begin{equation}
\Delta\alpha_{1}^{2}=\alpha^{2}-\alpha_{n1}^{2},\quad\Delta\alpha_{2}^{2}=\alpha^{2}-\alpha_{n2}^{2},
\end{equation}
where $\alpha_{n1}^{2}$ and $\alpha_{n2}^{2}$ are given by Eqs.~(\ref{eq:III_neutral_1})
and (\ref{eq:III_neutral_2}). As mentioned in Sec.~\ref{sec:section3}, we have $0\leq\alpha_{n1}^{2}<\alpha_{n2}^{2}\leq q^{2}$;
hence, $\Delta\alpha_{1}^{2}>\Delta\alpha_{2}^{2}$. If $\Delta\alpha_{1}^{2}>0$,
then $G(\Delta\alpha_{1}^{2})=|\Delta\alpha_{1}^{2}|H(-\Delta\alpha_{1}^{2})=0$,
hence $G(\Delta\alpha_{2}^{2})\geq G(\Delta\alpha_{1}^{2})$; if $\Delta\alpha_{1}^{2}<0$,
then $\Delta\alpha_{2}^{2}<\Delta\alpha_{1}^{2}<0$, hence we still
have $G(\Delta\alpha_{2}^{2})\geq G(\Delta\alpha_{1}^{2})$. This
shows that $G(\Delta\alpha_{2}^{2})\geq G(\Delta\alpha_{1}^{2})$
everywhere, and thus $\gamma_{2}\geq\gamma_{1}$ everywhere too. Correspondingly,
the largest TI growth rate is given by 
\begin{equation}
\gamma_{{\rm TI}}=|k_{x}u_{0}\Delta\alpha_{2}^{2}|q^{-2}\,\sqrt{1-\varrho^{-2}}
\end{equation}
at $\Delta\alpha_{2}^{2}<0$, and otherwise $\gamma_{{\rm TI}}=0$.
Accordingly, the TI develops when
\begin{gather}
\varrho^{2}>1,\quad\alpha^{2}<q^{2}-\bar{q}^{2}.
\end{gather}

Finally, since $|\Delta\alpha_{2}^{2}|=|\alpha^{2}-\alpha_{n2}^{2}|=q^{2}-\bar{q}^{2}-\alpha^{2}$
(assuming $\Delta\alpha_{2}^{2}<0$), the largest $\gamma_{{\rm TI}}$
is realized at $\bar{q}=0$; thus, $\phi(y)=\psi(y)$ {[}Eq.~(\ref{eq:III_psi_definition}){]}.
Then, using Eq.~(\ref{eq:III_perturbed_psi}) with $\bar{q}=0$ and
$n\to n_{2}$, we obtain the characteristic wavenumber of $\psi(y)$
as
\begin{equation}
|k_{y}|\doteq\sqrt{-\psi''/\psi}\approx\sqrt{-\Delta\alpha_{2}^{2}}=\sqrt{q^{2}-\alpha^{2}}<q.\label{eq:III_ky}
\end{equation}
Hence, the characteristic spatial scale of the TI mode is actually
larger than that of the ZF,  so the TI cannot be
described within the GO approximation in principle. Also note that
the same conclusion holds also for the barotropic vorticity equation
\cite{Kuo1949}. The only difference is that in the latter case, $\alpha^{2}=k_{x}^{2}$;
then, $\alpha^{2}$ is allowed to be zero, so Eq.~(\ref{eq:III_ky})
is less stringent, namely, $|k_{y}|\le q$. 

\subsection{There is no instability at $\boldsymbol{\alpha^{2}\geq q^{2}}$}

\label{subsec:section3.D}

Since $0\leq\alpha_{n1}^{2}<\alpha_{n2}^{2}$, the above results indicate
that there is no instability at $\alpha^{2}>\alpha_{n2}^{2}=q^{2}-\bar{q}^{2}$.
Strictly speaking, this conclusion is only valid when $\alpha^{2}$
is close to $\alpha_{n2}^{2}$. But one can also prove that no unstable
eigenmode exists if $\alpha^{2}\geq q^{2}$. Our proof follows the
argument from Ref.~\cite{Drazin1966}. The difference is that we
extend that argument by allowing for nonzero $\beta$, which is done
as follows.

Let us multiply Eq.~(\ref{eq:II_barotropic_vorticity}) by $\phi^{*}$
and integrate the result over $y$ from $0$ to $L$. After integrating
by parts, we obtain
\begin{multline}
\int_{0}^{L}|\phi'|^{2}dy+\alpha^{2}\int_{0}^{L}|\phi|^{2}dy\\
=-\int_{0}^{L}\frac{(U''-\beta)(U-C^{*})}{|U-C|^{2}}|\phi|^{2}dy.\label{eq:III_alphalimit_1}
\end{multline}
Let us multiply Eq.~(\ref{eq:III_RK}) by $C-C_{n}$ and add the
result to Eq.~(\ref{eq:III_alphalimit_1}). This gives
\begin{multline}
\int_{0}^{L}|\phi'|^{2}dy+\alpha^{2}\int_{0}^{L}|\phi|^{2}dy\\
=-\int_{0}^{L}\,\frac{(U''-\beta)(U+C_{n}-C-C^{*})}{|U-C|^{2}}\,|\phi|^{2}dy\\
=-\int_{0}^{L}\,\frac{(U''-\beta)}{(U-C_{n})}\,\frac{(U-C_{n})(U+C_{n}-2C_{r})}{|U-C|^{2}}\,|\phi|^{2}dy\\
=-\int_{0}^{L}\,\frac{(U''-\beta)}{(U-C_{n})}\,\frac{(U-C_{r})^{2}-(C_{n}-C_{r})^{2}}{|U-C|^{2}}\,|\phi|^{2}dy,
\end{multline}
where $C_{r}\doteq\text{Re}\,C$. Since $U=u_{0}\cos qy$ and $C_{n}=-\beta/q^{2}$,
one finds that $-(U''-\beta)/(U-C_{n})=q^{2}$ is a positive constant.
Hence,
\begin{multline}
\int_{0}^{L}|\phi'|^{2}dy+\alpha^{2}\int_{0}^{L}|\phi|^{2}dy\\
=q^{2}\int_{0}^{L}\frac{(U-C_{r})^{2}-(C_{n}-C_{r})^{2}}{|U-C|^{2}}|\phi|^{2}dy\\
<q^{2}\int_{0}^{L}|\phi|^{2}dy,
\end{multline}
so we obtain
\begin{equation}
\alpha^{2}<q^{2}-\frac{\int_{0}^{L}|\phi'|^{2}dy}{\int_{0}^{L}|\phi|^{2}dy}\leq q^{2}.
\end{equation}

This shows that no unstable eigenmode is possible if $\alpha^{2}\geq q^{2}$.
In application to the gHME, where $\alpha^{2}\equiv1+k_{x}^{2}$,
this indicates that not only do we have $|k_{y}|<q$, but the ZF itself
must also have $q^{2}\geq\alpha^{2}>1$ for the TI onset. In other
words, a ZF is $\textit{always}$ stable in the GO limit, because
in this limit we have $q^{2}\ll1$, and thus $\alpha^{2}>q^{2}$ automatically
(see also Sec.~\ref{sec:section5}).

\section{Comparison with numerical calculations}

\label{sec:section4}

Here, we compare our approximate analytic result {[}Eq.~(\ref{eq:III_Im(C)}){]}
with numerical solutions of Eq.~(\ref{eq:II_barotropic_vorticity}).
Specifically, we search for $\phi$ in the form of a Floquet mode,
\begin{eqnarray}
\phi & = & \sum_{n=-N}^{+N}\phi_{n}e^{i(\bar{q}+nq)y},
\end{eqnarray}
where the series has been truncated at a large enough $n=N$. Then,
the eigenvalues of Eq.~(\ref{eq:II_eigen}) are found numerically.

\begin{figure}
\includegraphics[width=1\columnwidth]{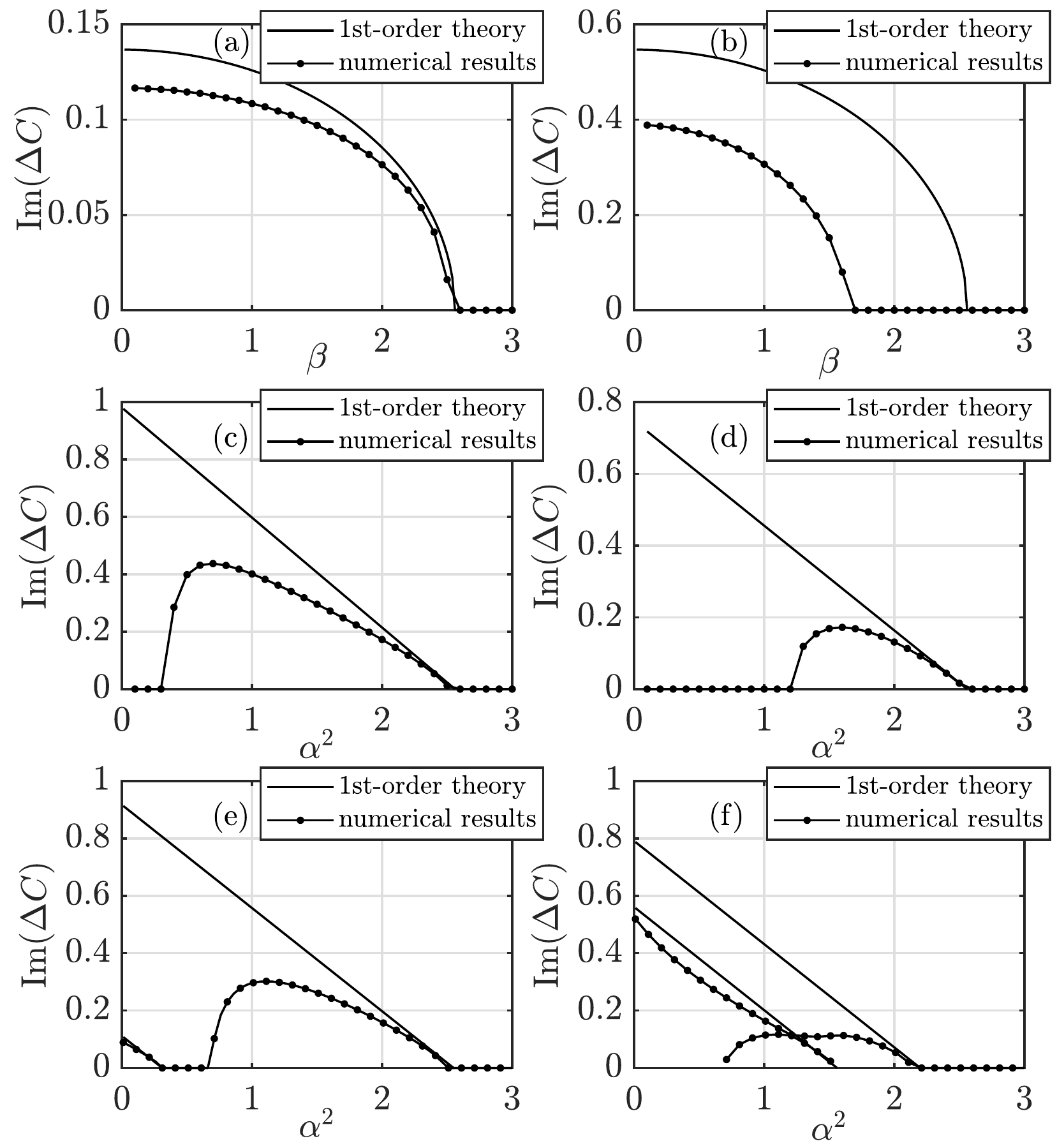}

\caption{(a)-(b): $\text{Im}(\Delta C)$ as a function of $\beta$ for: (a)
$\alpha^{2}=2.21$ and (b) $\alpha^{2}=1.16$. (c)-(d): $\text{Im}(\Delta C)$
as a function of $\alpha^{2}$ for: (c) $\beta=0.5$ and (d) $\beta=1.7$.
(e)-(f): $\text{Im}(\Delta C)$ as a function of $\alpha^{2}$ for:
(e) $\bar{q}=0.1$ and (f) $\bar{q}=0.6$. For (a)-(d), the parameters
are $q=1.6$, $\bar{q}=0$, and $u_{0}=1$. For (e)-(f), the parameters
are $q=1.6$, $\beta=1$, and $u_{0}=1$. The solid curves are calculated
using Eq.~(\ref{eq:III_Im(C)}), while the dotted curves are numerical
results obtained by solving Eq.~(\ref{eq:II_eigen}) with $N=50$.
Only positive $\text{Im}(\Delta C)$ are shown, but the complete plots
are symmetric with respect to the horizontal axis. This is because
if $C$ is a solution, then so is $C^{*}$ (Sec.~\ref{subsec:section2B}). [Associated dataset available at: \url{http://dx.doi.org/10.5281/zenodo.1241546}.] \cite{zenodo}
\label{fig:IV_1}}
\end{figure}

Figures \ref{fig:IV_1}(a) and (b) show the dependence of $\text{Im}(\Delta C)$
on $\beta$. It is seen that the perturbation theory {[}Eq.~(\ref{eq:III_Im(C)}){]}
approximates the numerical result with reasonable accuracy in Fig.~\ref{fig:IV_1}(a).
The discrepancy in Fig.~\ref{fig:IV_1}(b) is due to the fact that
the perturbation theory assumes $\alpha^{2}\approx\alpha_{n1,2}^{2}$,
which is not the case here. Also, in Fig.~\ref{fig:IV_1}(b) the
instability already vanishes at $\beta=1.6<q^{2}u_{0}=2.56$, which
indicates that $\varrho>1$ is only a necessary condition. 

Next, we consider the change of $\text{Im}(\Delta C)$ with $\alpha^{2}$
in Figs.~\ref{fig:IV_1}(c) and (d). Since $\bar{q}=0$, the two
neutral eigenmodes are at $\alpha_{n1}^{2}=0$ and $\alpha_{n2}^{2}=q^{2}=2.56$.
We also see that there are no unstable eigenvalues at $\Delta\alpha^{2}>0$,
which agrees with the perturbation theory. Unstable eigenvalues exist
when $\alpha^{2}\lesssim\alpha_{n2}^{2}=q^{2}$, whose numerical values
agree with perturbation theory when $\alpha^{2}\approx\alpha_{n2}^{2}$.
However, the range of $\alpha^{2}$ where $\text{Im}(\Delta C)>0$
depends on $\beta$; namely, the range is smaller when $\beta$ is
larger.

Finally, let us consider the dependence on $\bar{q}$. In Figs.~\ref{fig:IV_1}(e)
and (f), we plot the dependence of $\text{Im}(\Delta C)$ on $\alpha^{2}$
for nonzero $\bar{q}$. Then, $\alpha_{n1}^{2}=q^{2}-\bar{q}^{2}<q^{2}$
and $\alpha_{n2}^{2}=q^{2}-(q-\bar{q})^{2}>0$. In Fig.~\ref{fig:IV_1}(e),
$\bar{q}=0.1$ is small, and the existence of $\text{Im}(\Delta C)>0$
agrees with perturbation theory. In Fig.~\ref{fig:IV_1}(f), $\bar{q}=0.6$
is large, so $\alpha_{n1}^{2}$ and $\alpha_{n2}^{2}$ are close to
each other. Thus, the branch of unstable eigenvalues starting from
$\alpha_{n2}^{2}$ intersects with the branch starting from $\alpha_{n1}^{2}$. 

In summary, our first-order perturbation theory agrees with numerical
results when $\alpha^{2}$ is close to $\alpha_{n1,2}^{2}$. It also
captures the qualitative dependence on $\beta$. However, the dependence
on $\alpha^{2}$ away from $\alpha_{n1,2}^{2}$ is not well described.
An alternative approximation for $\gamma_{{\rm TI}}$, which is not
asymptotically accurate but applicable at all $\alpha^{2}$, was proposed
in Ref.~\cite{Zhu18a} based on a different argument.

\section{Tertiary instability in plasmas}

\label{sec:section5}

\begin{figure}

\includegraphics[width=1\columnwidth]{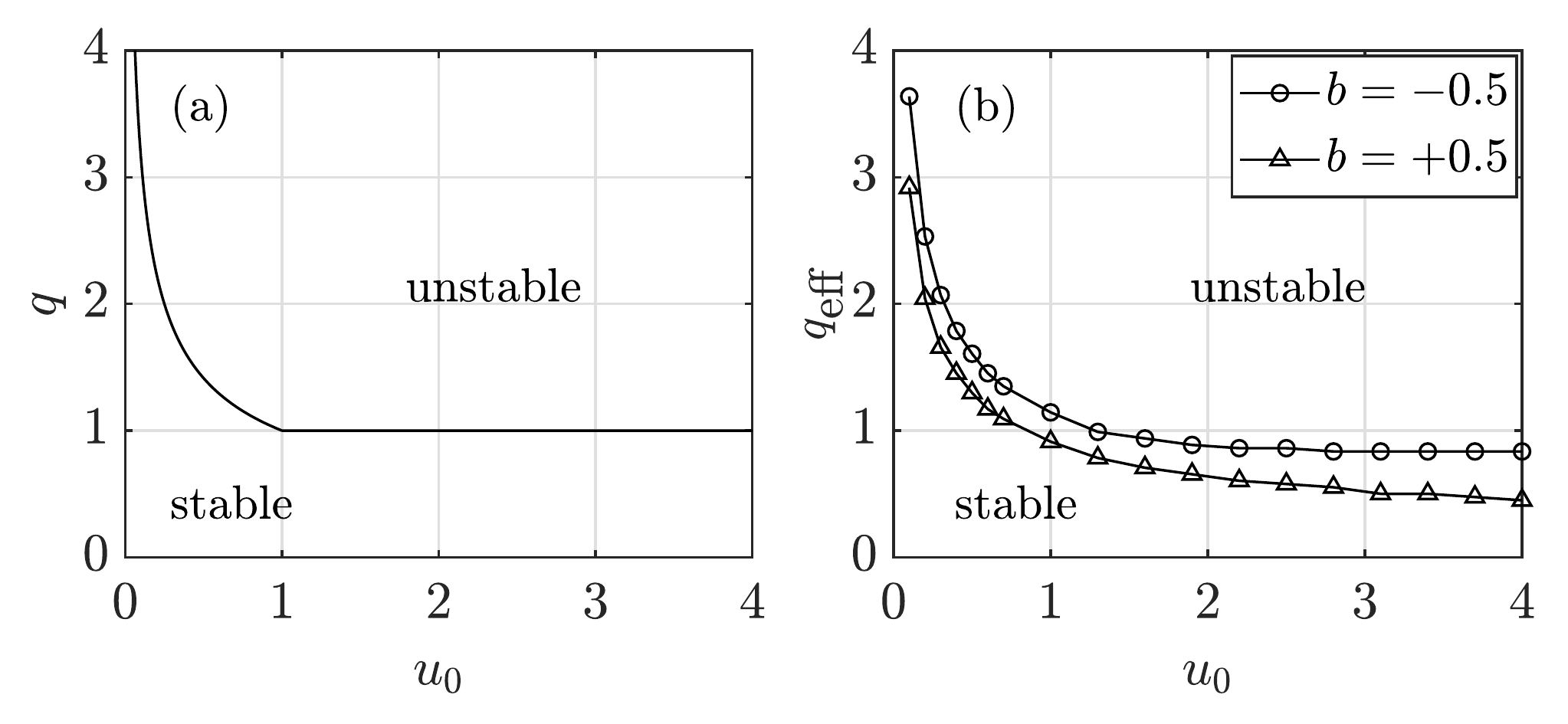}\caption{(a) The analytically calculated TI-stability diagram {[}Eq.~(\ref{eq:V_min_q}){]}
for a sinusoidal ZF within the gHME model. (b) A similar diagram found
numerically for non-sinusoidal ZFs given by Eq. (\ref{eq:VI_b-profile})
with $b=\pm0.5$, so $f_{{\rm eff}}=2.2$ (see Fig.~\ref{fig:VI_b-shape-fac}).
In both cases, $\beta=1$. [Associated dataset available at: \url{http://dx.doi.org/10.5281/zenodo.1241546}]. \cite{zenodo} \label{fig:V_separatrix}}
\end{figure}

Here, we consider the ramifications of our theory that are specific
to the plasma problem governed by the gHME as opposed to Kuo's geophysics
problem. The difference is in the definition of $\alpha^{2}$; specifically,
in the gHME, one has $\alpha^{2}=1+k_{x}^{2}$, so $\alpha^{2}>1$.
Hence, the results are as follows. The two necessary conditions for
the TI are (i) $q^{2}u_{0}>\beta$ and (ii) $q^{2}>1$. The second
condition comes from the fact that if $q^{2}\leq1$, then $\alpha^{2}\equiv1+k_{x}^{2}\geq q^{2}$
for any real $k_{x}$, and hence no instability is possible. We also
emphasize that the two necessary conditions combined together are
also $\textit{sufficient}$ for the TI. The reason for the sufficiency
is that when the two necessary conditions are satisfied, one can find
a nonzero $k_{x}$, such that $1+k_{x}^{2}$ is smaller but close
to $q^{2}$. Then, according to the analysis in Sec.~\ref{subsec:section3c},
there exists an unstable eigenmode. Hence, we conclude that for a
sinusoidal ZF, the necessary and sufficient condition for the onset
of TI is
\begin{gather}
q^{2}>q_{{\rm min}}^{2}\doteq\text{\ensuremath{\max}}\,\{\beta/u_{0},1\},\label{eq:V_min_q}
\end{gather}
as also illustrated in Fig.~\ref{fig:V_separatrix}(a).

As discussed in Sec.~\ref{subsec:section3c}, the TI growth rate
$\gamma_{{\rm TI}}\doteq|k_{x}\text{Im}(\Delta C)$| is found to be
\begin{gather}
\gamma_{{\rm TI}}=|k_{x}u_{0}|\vartheta H(\vartheta)\sqrt{1-\varrho^{-2}}.\label{eq:V_1st_growthrate}
\end{gather}
Here, $H$ is the Heaviside step function, $\vartheta\doteq1-(\bar{q}^{2}+1+k_{x}^{2})/q^{2}$,
and $\varrho=u_{0}q^{2}/\beta$. In a dimensional form, the two necessary
conditions found above are (i) $q^{2}>\rho_{s}^{-2}$ and (ii) $q^{2}u_{0}>\rho_{s}^{-2}V_{*}$,
and $\gamma_{{\rm TI}}$ is still given by Eq.~(\ref{eq:V_1st_growthrate}),
where
\begin{gather}
\vartheta=1-(\bar{q}^{2}+\rho_{s}^{-2}+k_{x}^{2})/q^{2},\quad\varrho=q^{2}\rho_{s}^{2}u_{0}/V_{*}.\label{eq:V_1st_growthrate_dimensional}
\end{gather}
Equations (\ref{eq:V_1st_growthrate}) and (\ref{eq:V_1st_growthrate_dimensional})
are among the main results of our paper.

As a corollary, the WKE, which assumes the GO limit that relies on
the assumption $q^{2}\ll\rho_{s}^{-2}$, is not adequate to describe
the TI. This conclusion applies to both the traditional WKE \cite{Diamond2005review,Smolyakov2000a,Smolyakov2000b,Malkov2001,Miki2012,kim2002,Smolyakov1999,Li2018}
and the ``improved'' WKE proposed recently in Refs.~\cite{Parker2016,Ruiz2016}.
(The improved WKE too relies on the assumption that $q$ is small
compared to the characteristic DW wavelength.) An adequate theory
of the TI must not assume the GO approximation. This is also discussed
in Ref.~\cite{Zhu18a}, where an alternative (but similar) approximation
for $\gamma_{{\rm TI}}$ is derived from different arguments.

Finally, let us compare our findings with those in other studies of
the TI. First, our findings support the conjecture in Ref.~\cite{Numata2007}
regarding the relevance of the RK criterion, and our approximate formula
for $\gamma_{{\rm TI}}$ {[}Eq.~(\ref{eq:V_1st_growthrate}){]} is
in general agreement with the numerical results in that paper. The
second part of our instability criterion, $q^{2}>\rho_{s}^{-2}$,
is not mentioned in Ref.~\cite{Numata2007} explicitly but is satisfied
for the simulation parameters presented there. Our approach is more rigorous than the truncated-Floquet approach used in Refs.~\cite{kim2002,Rath18,Zhu18a,St-Onge2017} in terms of predicitng the onset of the instability. (A comparison between the growth rates obtained from the two approaches can be found in Ref.~\cite{Zhu18a}.) In particular, the parameter $\beta$ does
not enter the final dispersion relation in Ref.~\cite{kim2002}, so the RK criterion is missed
\cite{foot:compare}. The discussion of the ``generalized
KHI'' in Ref.~\cite{kim2002}, as mentioned before, also seems
irrelevant to the TI. This is because the generalized KHI is excited
by a homogeneous background, which makes it just another branch of
the ZI. In any case, as is shown above, the WKE cannot capture the
TI in principle, because the underlying GO approximation will always
lead to the ZF amplification rather than deterioration \cite{Zhu18b}.

Our results are also different from those in Refs.~\cite{Rogers2000,Rogers2005}
in the following sense. In those papers, the TI is considered as a
mode driven by the ion temperature gradient, which is absent in our
model. Also, the mode structure in Refs.~\cite{Rogers2000,Rogers2005}
is found to be localized where $U'=0$. In contrast, the TI considered
in our paper (as well as in Refs.~\cite{kim2002,Numata2007}) is
similar to the KHI. The reason is that when $\beta=0$, our basic
equation (\ref{eq:II_barotropic_vorticity}) can also be used to describe
the KHI, in which case the mode amplitude peaks at $U''=0$. Also,
at arbitrary $\beta$, our Eq.~(\ref{eq:V_1st_growthrate_dimensional})
shows that when $q$ is large enough, one has $\gamma_{{\rm TI}}\approx|k_{x}u_{0}|$, which is also characteristic of the KHI.

\section{Non-sinusoidal zonal flow}

\label{sec:section6}

\begin{figure}
\includegraphics[width=1\columnwidth]{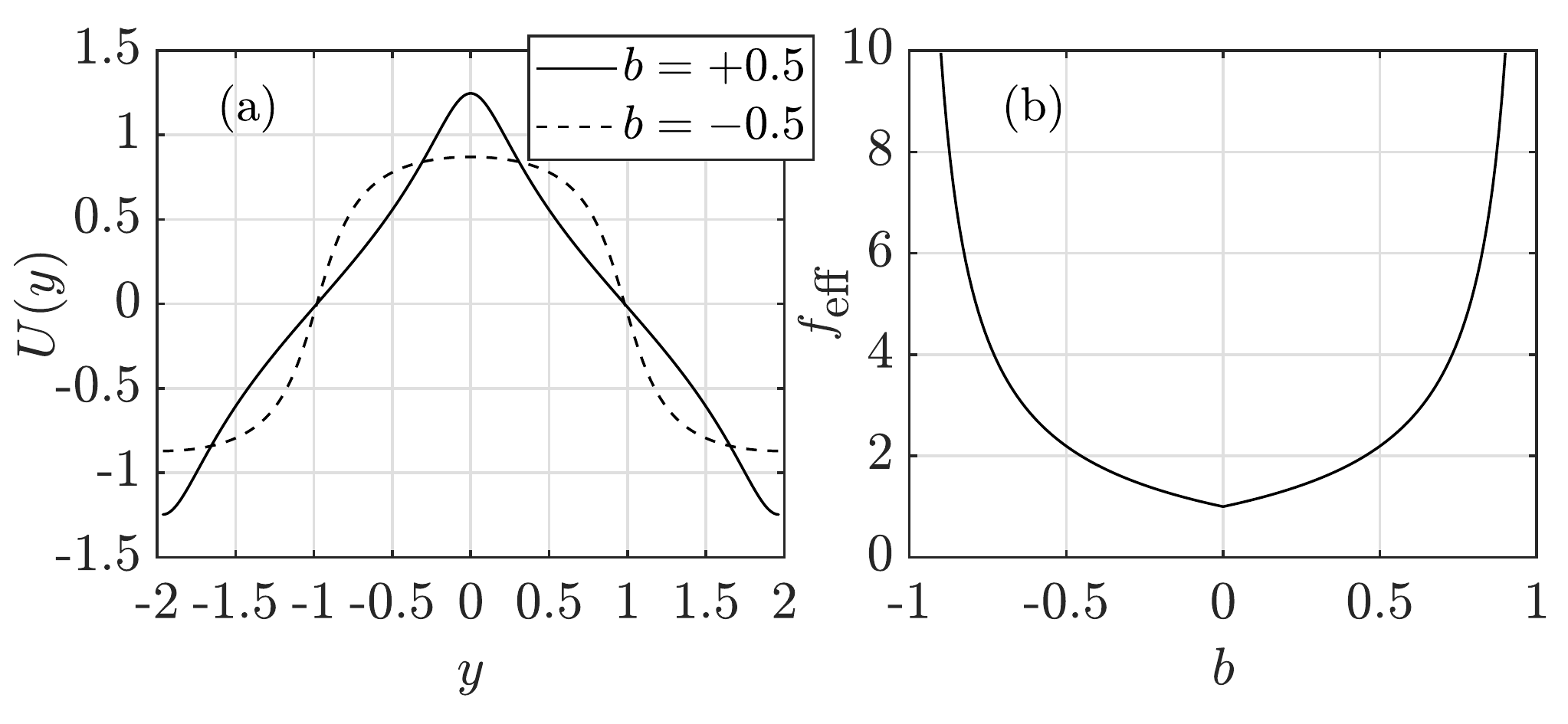}

\caption{(a) Two examples of non-sinusoidal ZFs [Eq.~(\ref{eq:VI_b-profile})]
with $b=\pm0.5$. (b) The factor $f_{\text{eff}}\protect\doteq q_{\text{eff}}/q$,
where the effective wavenumber $q_{{\rm eff}}$ is given by Eq.~(\ref{eq:VI_q_eff}).\label{fig:VI_b-shape-fac}}

\end{figure}

Finally, let us generalize our conclusions to nonsinusoidal flows
by considering a class of ZFs with a single control parameter $b$,
namely,
\begin{equation}
U(y)=u_{0}\sum_{k=0}^{+\infty}\,U_{k}\cos[(2k+1)qy],\label{eq:VI_b-profile}
\end{equation}
where $U_{k}=b^{k}/(2k+1)$ and $-1<b<1$ {[}Fig.~\ref{fig:VI_b-shape-fac}(a){]}.
When $b$ is zero, we recover Eq.~(\ref{eq:II_U_profile}). When
$b$ is nonzero, the ZF contains multiple harmonics, and in order
to describe its characteristic length scale, we define the effective
(weighted) wavenumber as
\begin{equation}
q_{\text{eff}}\doteq\sqrt{\frac{\sum_{k=0}^{+\infty}q_{k}^{2}|U_{k}|}{\sum_{k=0}^{+\infty}|U_{k}|}},\label{eq:VI_q_eff}
\end{equation}
where $q_{k}\doteq(2k+1)q$. We also define $f_{\text{eff}}\doteq q_{\text{eff}}/q$
{[}Fig.~\ref{fig:VI_b-shape-fac}(b){]}, which increases as $|b|$
increases. 

\begin{figure}
\includegraphics[width=1\columnwidth]{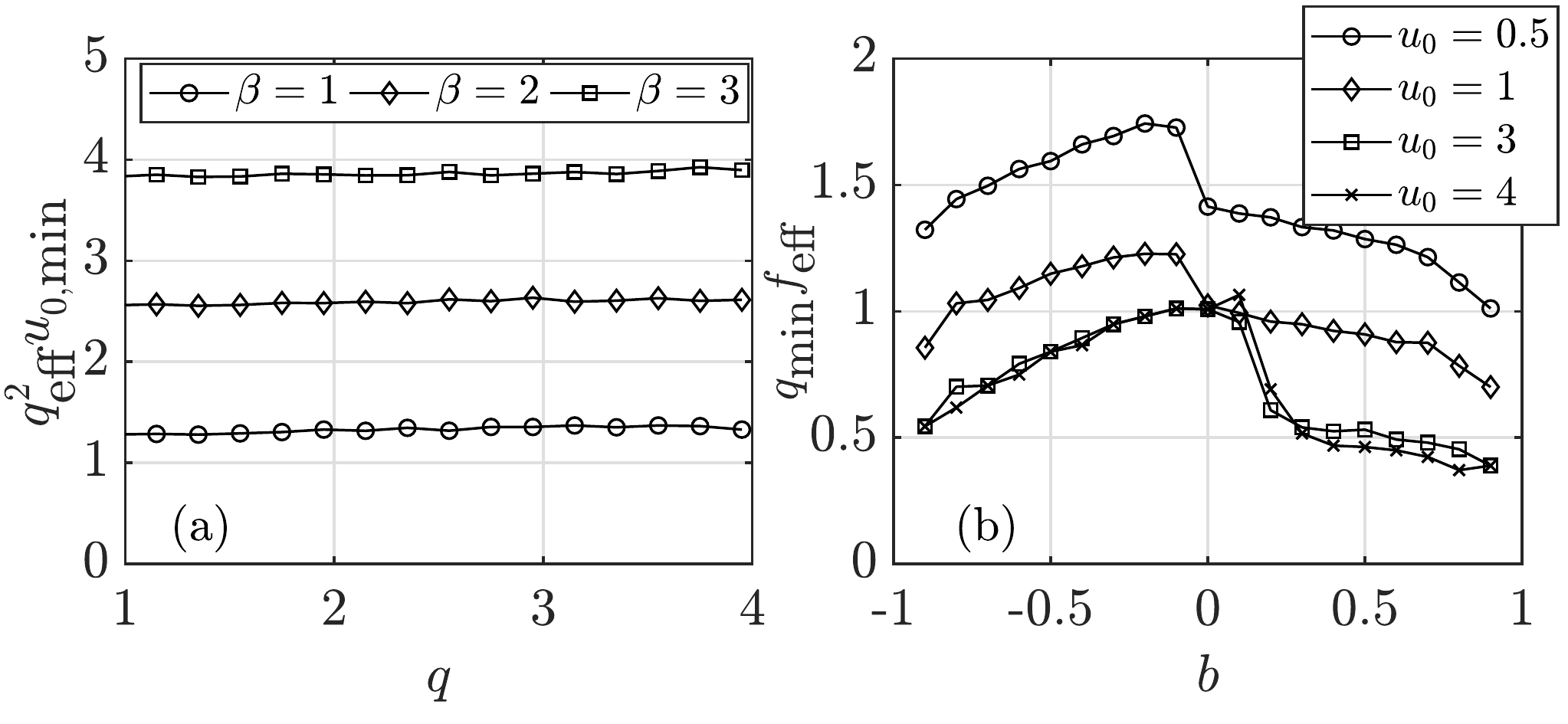}
\caption{(a) $q_{\text{eff}}^{2}u_{0,\text{min}}$ versus $q$. The effective
wavenumber $q_{\text{eff}}$ is given by Eq.~(\ref{eq:VI_q_eff}).
The ZF profile is given by Eq.~(\ref{eq:VI_b-profile}) with $b=-0.5$.
(b) $q_{\text{min}}f_{\text{eff}}$ versus $b$ for different $u_{0}$,
where $f_{\text{eff}}$ is given in Fig.~\ref{fig:VI_b-shape-fac}(b)
and $\beta=1$ is fixed. [Associated dataset available at: \url{http://dx.doi.org/10.5281/zenodo.1241546}.] \cite{zenodo}\label{fig:VI_two_conditions}}
\end{figure}

Knowing the non-sinusoidal ZF profile {[}Eq.~(\ref{eq:VI_b-profile}){]},
we can find numerical eigenvalues using the same procedure as in Sec.~\ref{sec:section4}.
First, we consider a fixed $q$ and numerically search for minimum
$u_{0}$ (denoted by $u_{0,\text{min}}$) below which unstable eigenvalues
disappear for all $k_{x}$. For example, Fig.~\ref{fig:VI_two_conditions}(a)
shows the results for $b=-0.5$. It is seen that $\varrho_{{\rm eff,min}}\doteq q_{{\rm eff}}^{2}u_{0,{\rm min}}/\beta$
remains approximately constant and close to one, namely, $\varrho_{{\rm eff,min}}\approx1.2$.
Similar results are obtained for other choices of $b$ within the
range $|b|\le0.9$. Thus, we conclude that $\varrho_{{\rm eff}}\doteq q_{{\rm eff}}^{2}u_{0}/\beta$
can be considered as an effective RK parameter for non-sinusoidal
ZFs; i.e., the first TI criterion holds approximately in the form
\begin{gather}
\varrho_{{\rm eff}}\apprge1.\label{eq:VI_first}
\end{gather}

Next, we consider a fixed $b$ and numerically search for the minimum
$q$ (denoted by $q_{\text{min}}$) below which unstable eigenvalues
disappear for all $k_{x}$. We find that for nonzero $b$, $q_{\text{min}}$
is smaller than one, and $q_{\text{min}}$ approaches zero as $|b|$
approaches one. However, $q_{{\rm eff,min}}\doteq f_{{\rm eff}}q_{{\rm min}}$
remains of order one. An example with $\beta=1$ is given in Fig\@.~\ref{fig:VI_two_conditions}(b),
where we show different results depending on the choice of $u_{0}$.
In particular, consider $b=0$, which corresponds to a sinusoidal
ZF (hence $f_{{\rm eff}}=1$). Then, for $u_{0}=0.5<1$, Eq.~(\ref{eq:V_min_q})
gives $q_{{\rm min}}=\sqrt{2}\approx1.4$, whereas for $u_{0}\ge1$,
one obtains $q_{{\rm min}}=1$. Also, the $u_{0}$-dependence becomes
weak at $u_{0}\geq3$. The results show that $q_{{\rm eff,min}}$
decreases slowly at nonzero $b$, and $q_{{\rm eff},\text{min}}\gtrsim0.5$
within the range $|b|\leq0.9$. Therefore, we conclude that the second
necessary condition for the TI still holds approximately for non-sinusoidal
ZFs, namely, in the form
\begin{gather}
q_{{\rm eff}}\apprge1.\label{eq:VI_second}
\end{gather}

Finally, the numerically-found stability diagram for nonzero $b$ is shown in Fig.~\ref{fig:V_separatrix}(b).
It is seen that the diagram is very similar to the sinusoidal case.
Therefore, just like for sinusoidal ZFs, the two necessary conditions
combined together {[}i.e., Eqs.~(\ref{eq:VI_first}) and (\ref{eq:VI_second}){]}
are also sufficient for the TI.

\section{conclusions}

\label{sec:section7}

In summary, we explored the tertiary instability (TI) of zonal flows
within the gHME model. Our analytic calculation extends and revises
Kuo's analysis of the mathematically similar barotropic vorticity
equation for incompressible neutral fluids on a rotating sphere \cite{Kuo1949};
then, the results are applied to the plasma case. An error in Kuo's
original results is pointed out. An explicit analytic formula for
the TI growth rate $\gamma_{{\rm TI}}$ is derived {[}Eq.~(\ref{eq:III_Im(C)})
and (\ref{eq:V_1st_growthrate_dimensional}){]} and compared with
numerical calculations. It is shown that, within the generalized Hasegawa--Mima model, a sinusoidal ZF is TI-unstable if and only if it satisfies the Rayleigh--Kuo criterion (known from geophysics) and that the ZF wave number exceeds the inverse ion sound radius. For non-sinusoidal ZFs, the results
are qualitatively similar. As a corollary, there is no TI in the
GO limit, i.e., when the perturbation wavelength is small compared to the ZF
scale. This also means that the traditional wave kinetic equation, which is derived
under the GO assumption, cannot adequately describe the ZF stability.
\begin{acknowledgments}
This work was supported by the U.S. Department of Energy (DOE), Office
of Science, Office of Basic Energy Sciences, and also by the U.S.
DOE through Contract No. DE-AC02-09CH11466.
\end{acknowledgments}


\begin{thebibliography}{10}
\bibitem{foot:coordinate}Equation (\ref{eq:II_barotropic_vorticity})
is also identical to Eq.~(18) in Ref.~\cite{Numata2007} except
that $x$ and $y$ are swapped because the notations are different.
\bibitem{foot:enstrophy}The gHME conserves the total enstrophy. Hence,
provided that the initial conditions are physical (smooth enough),
infinite-enstrophy modes cannot be excited.
\bibitem{foot:viscosity}As noted in Sec.~\ref{subsec:section2B},
this is not the case in the presence of viscosity. Nevertheless, this
does not reinstate Kuo's argument, because the latter was developed
within a model of a non-viscous fluid.
\end{thebibliography}
\end{document}